# Gas temperature effect on the time for onset of particle nucleation in argon diluted acetylene plasma

I. Stefanović[1,2], E. Kovačević[1], J. Berndt[1], and J. Winter[1]

[1]*Institute for Experimental Physics II, Ruhr-University, 44780 Bochum, Germany*
[2]*Also at Institute of Physics, POB 57, 11000 Belgrade, Serbia and Montenegro*

## 1. Introduction

Particle nucleation in processing plasmas is an important issue that was studied both experimentally and theoretically. A better understanding of particle nucleation may help us to understand the basic problems of the involved plasma chemistry. On the other hand it is interesting to be able to control the production of the particles.

One of the observed phenomena is a delay of particle nucleation in argon-silane plasmas due to an increase of the gas temperature [1, 2]. It is one of various "critical phenomena" that has been recognized in the process of dust formation. It was recently proposed [3] that the temperature dependence of the Brownian diffusion is the most important effect for the nucleation delay in argon-silane plasmas.

In our work we focused on the study of powder formation in $C_2H_2/Ar$ plasmas according to the fact that acetylene as monomer plays an important role in rapid powder formation. In this scope we used a combination of FTIR and mass spectroscopy, which are the mostly used experimental techniques for the diagnostics of powder formation. To test the proposed mechanism for particle nucleation delay we measured the particle nucleation under different plasma conditions: firstly we increased the gas temperature and secondly we changed the background gas from argon to helium.

## 2. Experiment

Our experimental set-up consists of a parallel-plane electrode system, 30 cm in diameter, forming a 8 cm wide gap. The electrodes are symmetrically capacitively coupled on a 13.6 MHz power supply witch is used for standard industry applications. The electrode system is confined in a stainless-steel vacuum vessel, supplied with several observation ports which are used for: Fourier – Transformed Infrared absorption (FTIR), mass spectroscopy, plasma emission spectroscopy, Langmuir probe measurements, and dust collection. An argon acetylene or argon/ helium gas mixture in the ratio of 8 sccm (standard cubic centimetre pro minute) to 0.5 sccm was continuously supplied to the discharge chamber by means of gas flow controller. More details of our experimental set-up and procedure can be found in our previous papers [4, 5].

## 3. Results

During the operation of the plasma we observed a periodically repeating growth and disappearance of the dust particles with a time period of approximately 25 min. This periodical behaviour was detected by means of laser and/or IR extinction measurements. We explained this effect by means of a simple model [4]: after the nucleation, the dust particles continue to grow by accretion. The balance of the different forces acting on the particles (ion drag, electrostatic, gravitational, neutral drag, and thermophoresis) keep the dust levitating inside the plasma (for our experimental conditions the dust is not only present in the vicinity of the plasma sheets but also in the plasma bulk). The diffusion to the walls/electrodes is constricted through the plasma potential. According to the different dependence upon the particle radius, the balance of these forces is broken after the dust has reached a certain radius. The dust then disappears from the plasma and the new growth process starts. Before the dust disappears we observed the formation of s void in the middle between the electrodes.

## 3.1. Temperature dependence

In our experiment we were able to increase the gas temperature by heating the stainless-steel vacuum vessel. We worked at two different temperatures: at room temperature (27°C ) and at a temperature of 80°C.

Fig. 1 shows the temporal development of the IR signal obtained from measured FTIR spectra [4] for two different gas temperatures $T = 27°C$ and $T = 80°C$ in a argon/acetylene discharge. It can be seen that in the case of the elevated gas temperature the particle nucleation/coagulation starts several minutes later for the first cycle. The time between two consecutive cycles is also in the same order of magnitude larger.

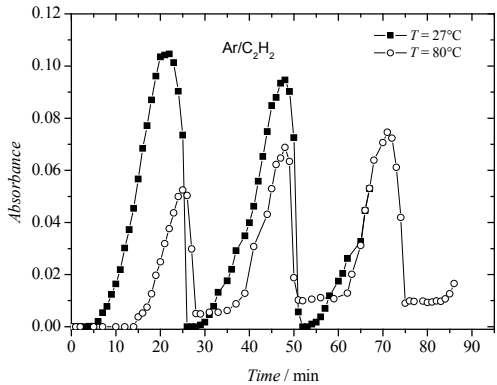

Fig. 1: Temporal development of extinction signal of dust particles for two different gas temperatures - solid squares- $T = 27°C$, open circles - $T = 80°C$.

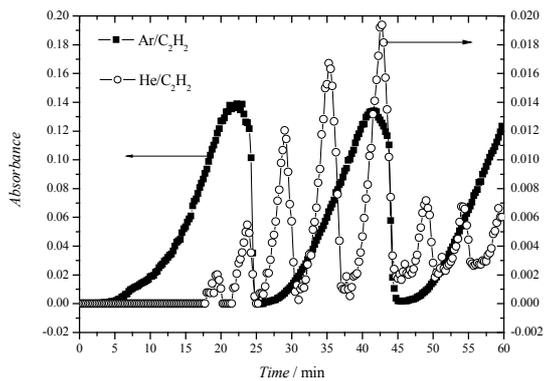

Fig. 2: Temporal development of extinction - scattering signal of dust particles for two different background gases - solid squares- argon, open circles - helium.

## 3.2 The influence of the background gas

To test the influence of the background gas on the particle nucleation delay we changed from the argon/acetylene to the argon/helium mixture, keeping the gas flow rates identical. The results are presented on Fig. 2. The time for starting the first growth cycle in the He/$C_2H_2$ mixture is drastically increased whereas the period time of the observed oscillation is decreased. The changing of the oscillation amplitude of the IR signal in the case of the He/$C_2H_2$ discharge can be attributed to a changing of the plasma conditions in the first few cycles. The impurities in the plasma chamber (oxygen and water) can change the plasma conditions and the capability of the plasma to carry larger dust particles. After a few cycles, the chamber walls and electrodes are cleaned or covered with a hydrocarbon film, thus eliminating the main source of impurities [4]. After this time the oscillation amplitude stays almost constant.

## 4. Discussion

In their recent paper Bhandakhar, Kortshagen and Grishick used a quasi-one-dimensional model to simulate particle generation in argon-silane plasma [3]. They performed a comparative study of all explanations for nucleation delay proposed: the temperature dependence of the de-excitation of vibration ally excited species, the temperature dependence of the electron attachment rate, gas density effects, and the temperature dependence of the Brownian diffusion coefficient. Only the latter effect could, at least quantitatively explain the observed delay in nucleation.

Our experimental results in different polymerisation systems support their conclusion because of different processes leading to the formation of the dust. The increase of the gas temperature leads to increase of nucleation delay, according to Fig. 2. In the same manner, by switching from argon to helium as the background gas, we increased Brownian diffusion coefficient which governs the nucleation delay (see equation 4 in [3]).

5. Conclusion

In this paper we analysed the time delay for onset of nucleation in hydrocarbon plasma. The two different conditions were tested, which should be critical for the observed nucleation delay: gas temperature increase and background gas effect. In both cases the increase of the delay time was observed. This is in accordance with the results of the quasi-one –dimensional model [3] tested on silane plasma.